\documentclass[dvips]{arxstspdf}
\usepackage{flushend}

\volume{24}
\issue{4}
\pubyear{2009}
\firstpage{387}
\lastpage{387}
\doi{10.1214/09-STS310}

\begin{document}
\begin{frontmatter}

\title{Introduction to the Special Issue: Genome-Wide Association
Studies}
\runtitle{GWAS}

\begin{aug}
\author[a]{\fnms{Gang} \snm{Zheng}\corref{}\ead[label=e1]{zhengg@nhlbi.nih.gov}},
\author[b]{\fnms{Jonathan} \snm{Marchini}\ead[label=e2]{marchini@stats.ox.ac.uk}}
\and
\author[c]{\fnms{Nancy L.} \snm{Geller}\ead[label=e3]{gellern@nhlbi.nih.gov}}
\runauthor{G. Zheng, J. Marchini and N. L. Geller}

\affiliation{National Heart, Lung and Blood Institute, University of
Oxford and National Heart, Lung
and Blood Institute}

\address[a]{Gang Zheng is a Mathematical Statistician, Office of
Biostatistics Research, Division of Cardiovascular Diseases (DCVS),
National Heart, Lung and Blood Institute, 6701 Rockledge Drive,
Bethesda, Maryland
20892-7913, USA \printead{e1}.}
\address[b]{Jonathan Marchini is a University Lecturer in Statistical Genomics in
the Mathematical Genetics Group, Department of Statistics, University of
Oxford, 1 South Parks Road, Oxford OX1 3TG, UK \printead{e2}.}
\address[c]{Nancy Geller is the
Director of Office of Biostatistics Research, DCVS, National Heart, Lung
and Blood Institute, 6701 Rockledge Drive, Bethesda, Maryland 20892-7913, USA \printead{e3}.}
\end{aug}



\end{frontmatter}

Genome-wide association studies (GWAS) have recently provided some
exciting scientific advances for identifying susceptibility genes for
many common diseases (e.g., WTCCC, 2007, in the June issue of
\textit{Nature}). These advances will have great impact on future
genetic studies and clinical trials to understand mechanisms of common
diseases. The GWAS revolution is far from having run its course.
Attention is now turning from simple case-control designs to using
family data or prospective cohort data that convey a wealth of
phenotypic information, including life-course varying phenotypes, as
well as environmental exposures. Statistical analyses of GWAS data have
to date focused mainly on the simplest tests of one SNP at a time,
leaving open the possibility that more sophisticated analyses may reveal
further important results. These could involve results about metabolic
pathways, gene-by-gene and gene-by-environment interactions, imputations
and copy number variants. Funds from the US government (National
Institutes of Health), the Wellcome Trust and other research
organizations for genomic studies have increased substantially.

Advances in statistical methodology will be central in these
developments, and we proposed this special issue to help foster them, by
authoritative reviews of the current state of the art, and pointers to
novel developments. Statistical issues and challenges arise from all
aspects of design and analysis of GWAS. This special issue consists of
12 papers focusing on the following topics: statistical designs using
case-control or family data, two and multi-stage designs for GWAS,
retrospective and prospective designs and their impact on single marker,
imputation and haplotype analyses, robust single marker analysis and
probability measures of detecting markers with true associations,
population structure and how to detect and correct for it, multiple
testing issues and weighted hypothesis testing for GWAS, strategies for
analyses of gene--gene and\break gene-environmental interactions in the GWAS
setting, novel Bayesian approaches to impute and test markers,
estimating genetic effect and making predictions using significant
markers from GWAS, analysis of copy number variants, and replication
studies for GWAS. The authors of these reviews are widely published in
this area and we thank them for their timely contributions. Without
their enthusiasm this issue could not come to fruition.

Some other topics are not included in this special issue, including
genotype call algorithms and methods of quality control, integration of
gene expression and genetic data, random forests and machine learning
methods, and methods related to resequencing genetic data. An overall
review of Bayesian methods in genetic association studies, including
GWAS, a topic not included in this issue, has recently been published by
Stephens and Balding (2009) in the October issue of \textit{Nature
Reviews Genetics}. Much computer software is available to run analyses
for GWAS. A review of these computer programs is not presented in this
special issue, but a number of programs are already included in the
various reviews. Study designs combining case-control and family\break data
for GWAS are not included in this special issue.

Finally, we would like to thank Peter Donnelly and David Balding for
their support of the idea to have this special GWAS issue. In
particular, David made very helpful suggestions to our initial proposal
of this special issue. We would also like to thank the Executive Editor,
David Madigan, for his excellent editorial support. All published papers
in this special issue have been through the regular peer-review process.
Our thanks also go to the referees of this special issue.

\end{document}